\documentclass[useAMS,usenatbib]{mnras}

\usepackage[usenames,dvipsnames]{xcolor}

\usepackage{amssymb, amsmath}
\usepackage{epsf}
\usepackage{bm}

\usepackage{natbib}
\usepackage{graphicx}
\usepackage{cancel}

\def\la{\; \raise0.3ex\hbox{$<$\kern-0.75em\raise-1.1ex\hbox{$\sim$}}\;}
\def\ga{\;  \raise0.3ex\hbox{$>$\kern-0.75em\raise-1.1ex\hbox{$\sim$}}\;}

\title[Long-lasting accretion-powered chemical heating of millisecond pulsars.]
{Long-lasting accretion-powered chemical heating of millisecond pulsars}

\author[E. M. Kantor, M. E. Gusakov]
{E.~M.~Kantor\thanks{kantor@mail.ioffe.ru},
M. E. Gusakov\thanks{gusakov@astro.ioffe.ru},
\\
$^1$Ioffe Institute,
Polytekhnicheskaya 26, 194021 St.-Petersburg, Russia
}

\begin{document}

\date{Accepted 2021 xxxx. Received 2021 xxxx;
in original form 2021 xxxx}

\pagerange{\pageref{firstpage}--\pageref{lastpage}} \pubyear{2021}

\maketitle

\label{firstpage}

%
\begin{abstract}
We analyze the effect of 
magnetic field in 
superconducting neutron-star cores
	on the chemical heating of millisecond pulsars (MSPs). 
	We argue that the magnetic field destroys proton superconductivity in some volume fraction 
	of the stellar core,
thus allowing
	for unsuppressed non-equilibrium reactions 
	of particle mutual transformations there. The reactions transform the chemical energy, 
	accumulated by a neutron star 
	core 
during the low-mass X-ray binary stage, into heat. 
	This heating may keep
	an NS warm at the MSP stage
	(with the surface temperature $\sim 10^5\,\rm K$) 
	for more than a billion of years after ceasing of accretion,
	without appealing to the rotochemical heating mechanism.
\end{abstract}
%

\begin{keywords}
stars: neutron -- stars: interiors -- pulsars -- nuclear reactions -- ultraviolet: stars
\end{keywords}

\maketitle

\section{Introduction}

%
Rather high values of surface temperatures 
measured for two millisecond pulsars (MSPs), 
PSR J0437-4715 (\citealt{kargaltsev04,durant12,bogdanov19}) 
and PSR J2124-3358 (\citealt{rangelov17}),
imply that some reheating mechanism 
should operate in these objects.
A number of such mechanisms exists in the literature, 
the most promising 
are 
vortex creep (\citealt*{alpar84}), 
rotochemical heating in the crust (\citealt*{gkr15}), 
and rotochemical heating in the core (\citealt*{reisenegger95}).
Here we focus on the latter mechanism.
Calculations by
\cite*{pr10} and \cite*{reisenegger15} 
suggest that rotochemical heating can likely explain the observed temperatures of
PSR~J0437$-$4715 and PSR J2124-3358 if
either the proton or neutron superfluid energy gaps 
(or both)
are sufficiently large
in the whole stellar core.
However, 
the available
analysis of the rotochemical heating 
did not account for the possible 
presence of the magnetic field in the core.
Moreover, the effect of the neutron-star (NS) matter compression during the preceding accretion at the low-mass X-ray binary (LMXB) stage 
has never been studied. 
Our aim here is to fill these gaps.

The paper is organized as follows.
Section \ref{st} discusses available observations of MSP surface temperatures.
In Section \ref{approach} 
we present general equations that govern 
the combined NS thermal and chemical evolution
under compression.
The adopted physics input is described in Section \ref{input}.
In Section \ref{SC} we analyze the effect of the core magnetic field on the chemical heating.
Section \ref{acc} examines the effect of the 
accretion during the LMXB stage 
on the thermal states of MSPs.
Section \ref{assumptions} contains discussion of the sensitivity of our results 
to various approximations made in the paper.
We conclude in Section \ref{conc}.

\section{Observed surface temperatures of millisecond pulsars}
\label{st}

A millisecond pulsar with the best constraint on the surface temperature 
is PSR J0437-4715 (hereafter, `J0437'; \citealt{kargaltsev04,durant12}). 
The mass of J0437 is 
measured to be 
$M=(1.3-1.58)M_\odot$ 
with $3\sigma$ significance (\citealt{reardon16}).
The rotation 
rate 
and spin-down rate 
(accounting for the Shklovskii effect) 
equal, respectively, $\nu=173.7\,\rm Hz$ and $\dot{\nu} = -4.14\times 10^{-16}\,\rm Hz \,s^{-1}$. 
Effective redshifted (seen by a distant observer) surface temperature is estimated as $T^\infty_{\rm s}=(1.0-1.5)\times 10^5\,\rm K$ (with about a
$15\%$ uncertainty dominated by the uncertainty in the extinction)
for apparent, as seen by a distant observer, radius, $R_\infty=(18-13)\,\rm km$ (\citealt{durant12}). 
While \cite{durant12} obtained the surface temperature of J0437 
by interpreting its thermal spectrum as blackbody emission, 
\cite{ggr19} fitted the same spectrum with atmospheric models and found that the spectral fits favour a hydrogen atmosphere with the
surface redshifted temperature 
$T^\infty_{\rm s}=(2.3\pm 0.1)\times 10^5\,\rm K$ 
and stellar coordinate radius $R=13.6^{+0.9}_{-0.8}\,\rm km$. 
The lower/upper uncertainties provide 68\% credible intervals.
Later,
by fitting new observational data from NICER with the hydrogen atmosphere model, 
\cite{bogdanov19}
reported the local effective surface temperature 
$T_{\rm s}=(1.8^{+0.2}_{-0.6})\times 10^5\,\rm K$ 
and stellar coordinate radius $R=15.3^{+2.0}_{-1.6}\,\rm km$ (90\% credible intervals).

Another pulsar with the measured surface temperature is PSR J2124-3358. 
The latter is estimated to be $(0.5-2.1)\times 10^5\,\rm K$ 
assuming the NS radius, as seen by a distant observer,
$R_\infty=12\,\rm km$ (\citealt{rangelov17}). 
The rotation rate of PSR J2124-3358, $\nu=202.8\,\rm Hz$, and its spin-down energy loss rate, $\dot{E} = 6.8\times 10^{33}\,\rm erg \,s^{-1}$, make this pulsar very similar to J0437.

There is also a 
set of
upper limits on the surface temperatures of a number of 
millisecond pulsars (see, e.g., \citealt{schwenzer17,chugunov17,hhc19,boztepe19}),
which are, however, all rather high, being not very restrictive for
rotochemical heating theory.

In view of the above mentioned facts, 
in what follows we 
only confront	
our results 
with the observations of the pulsar J0437,
assuming that 
its effective redshifted surface temperature is 
$T^\infty_{\rm s}=(1.0-1.5)\times 10^5\,\rm K$, in accordance with \cite{durant12,bogdanov19}.

\section{General equations and our approach}
\label{approach}

To describe NS evolution under combined action of compression 
caused by
accretion and spin-down, 
we follow the approach developed by \cite{fr05}. 
Initially, this approach was
applied to study
departure from the beta-equilibrium of the 
spinning-down neutron star.
However, an NS may accrete about $\sim 0.1M_\odot$ 
during its evolution in LMXB (\citealt{ozel12,antoniadis16}). 
Such a substantial amount of the accreted material 
compresses NS matter and could lead to strong deviations from 
chemical equilibrium. 
Thus, here we generalize the framework of \cite{fr05} to account, in addition, for 
the compression of matter due to accretion from the low-mass companion in an LMXB.
Following \cite{fr05,rjfk06}, 
we assume that the redshifted chemical potentials $\mu_{\rm n}^\infty$, $\mu_{\rm \mu}^\infty+\mu_{\rm p}^\infty$, and $\mu_{\rm e}^\infty+\mu_{\rm p}^\infty$ (as seen by a distant observer) 
are uniform 
because of the
very efficient diffusion (\citealt{reisenegger97, dgs20}),
while the redshifted internal temperature, $T^\infty$, 
is uniform due to high thermal conductivity (\citealt{ykgh01}). 
Here and below the subscripts $\rm b,n,p,e,\mu$ 
refer to baryons, neutrons, protons, electrons, and muons, respectively. 
The temperature $T^\infty$ is driven by the thermal balance equation (\citealt{yls99,fr05})
\begin{eqnarray}
C\dot{T}^\infty=L^\infty_{\rm acc}-L^\infty_{\gamma}+\int_V \left(Q^\infty_{\rm heat}-Q^\infty_{\nu}\right)dV, \label{Teq}
\end{eqnarray}
where dot stands for the time derivative; 
$dV=4{\rm \pi}\,r^2\, {\rm e}^{\lambda/2}dr$ is the proper volume element, 
with ${\rm e}^{\lambda}$ being the
radial component of the metrics of a non-rotating reference star
(the effect of rotation on the metrics is assumed to be small and is neglected in this equation).
$C$ in equation (\ref{Teq}) is the heat capacity of an NS
and
$Q^\infty_{\nu}$ is the neutrino emissivity.
In what follows, we assume that the direct Urca (DUrca) processes 
(\citealt{gs41}) 
are closed (their effect is discussed in Section \ref{assumptions}) 
and account for the two main contributions to $Q^\infty_{\nu}$. 
The first one comes from the non-equilibrium modified Urca (MUrca) processes
in the core: 
\begin{eqnarray}
B_i+{\rm n} \rightarrow B_i+{\rm p}+{\rm l} + \overline{\nu}_{\rm l}, \;\;\;\; B_i+{\rm p}+{\rm l} \rightarrow B_i+{\rm n} + \nu_{\rm l}.  
\end{eqnarray}
Here $B_i=\rm n,p$  stands for, respectively, 
neutron and proton branches of MUrca reactions, $\rm l={\rm e,\mu}$.
The second contribution to $Q^\infty_{\nu}$ comes from the
baryon-baryon bremsstrahlung:
\begin{eqnarray}
B_i+B_k \rightarrow B_i+B_k + \nu+ \overline{\nu},
\label{brems}
\end{eqnarray}
where $B_i,B_k$ stands for $\rm n$ or $\rm p$; $\nu,\overline{\nu}$ are neutrino and antineutrino 
of one of the three possible flavors. 
Generally, baryon-baryon bremsstrahlung is much weaker than MUrca processes. 
Still, if superfluidity (superconductivity) of neutrons (protons) strongly suppresses 
MUrca 
reactions,
then bremsstrahlung with protons (neutrons) becomes the main contributor to $Q^\infty_{\nu}$. 
We emphasize, however, that the process (\ref{brems})
does not change the chemical composition of the stellar matter. 
Note that we do not account for the explicit contribution to $Q^\infty_\nu$ 
of the Cooper pairing neutrino emission process (e.g., \citealt{ykgh01}),
but discuss its possible role in Section \ref{assumptions}.
Further, $Q^\infty_{\rm heat}$ in equation (\ref{Teq}) represents 
the heat release
in the non-equilibrium reactions
\begin{eqnarray}
Q^\infty_{\rm heat}=\sum_{\rm l=e,\mu}{\eta_{\rm l}^\infty \Delta \Gamma_{\rm l}\, {\rm e}^{\nu/2}}, \label{heat}
\end{eqnarray}
where $\eta_{\rm l}^\infty\equiv \mu_{\rm n}^\infty-\mu_{\rm l}^\infty-\mu_{\rm p}^\infty$ 
is the redshifted imbalance of chemical potentials (uniform throughout the core and vanishing in equilibrium); 
$\Delta \Gamma_{\rm l}$ is the net 
production
rate 
of particle species $\rm l=\rm{e,\mu}$ due to reactions per unit volume (\citealt{ykgh01});
$-{\rm e}^{\nu}$ is the
time component of the metrics of a nonrotating reference star. 
 Further,
\begin{eqnarray}
L^\infty_{\gamma}=4{\rm \pi}\sigma R^2 T_{\rm s}^4 {\rm e}^{\nu_{\rm s}}
\end{eqnarray}
accounts for the photon emission from the surface. Here $R$ is the stellar
coordinate radius, $\nu_{\rm s}=\nu(R)$. 
We relate the effective surface temperature in quiescence,
$T_{\rm s}$,
to the internal temperature, $T$, using the fitting formula (A8) 
from \cite{pcy97}, relevant 
to fully accreted envelope. 

Finally, $L_{\rm acc}^\infty$ in equation (\ref{Teq}) is the heating rate due to accretion
\begin{eqnarray}
L_{\rm acc}^\infty\approx\frac{\dot{M}}{m_{\rm u}}\,q\,{\rm e}^{\nu_{\rm s}/2},
\end{eqnarray}
where ${m_{\rm u}}$ is the nucleon mass unit, $\dot{M}$ is the average accretion rate 
(mass accreted per unit time of a distant observer), 
$q$ is the heat released per one accreted baryon
because of non-equilibrium nuclear reactions
in the crust (deep crustal heating). 
We approximate the redshift in the crust 
by
the surface redshift, $\nu\approx \nu_{\rm s}$. 
The values of $\dot{M}$ and $q$ are rather uncertain. 
In our analysis we choose $\dot{M}=10^{-10}M_\odot/\rm yr$ 
($M_\odot$ is the solar mass) and $q=0.5\,\rm MeV/{\rm baryon}$, 
in accordance with \cite{gc20,gc21}, 
who found that $q\la 0.5\,\rm MeV/{\rm baryon}$. Note, however, that our results are not sensitive to the choice of these values (see Section \ref{SF}).

In addition to the thermal balance equation (\ref{Teq}), 
one should also 
formulate evolution equations
for the redshifted chemical potential imbalances, $\eta_{\rm l}^\infty$. 
To establish these equations, let us 
introduce the vectors $\pmb \eta\equiv (\delta \mu_{\rm n},\eta_{\rm e}, \eta_{\rm \mu})$ 
(where $\delta \mu_{\rm n}$ is the departure of the neutron chemical potential 
from its equilibrium value; 
we introduce $\delta \mu_{\rm n}$ purely for mathematical reasons, 
in order to invert the matrix $\textbf{\textsf{A}}_{\rm (\mu)}$
appearing below) 
and $\pmb {\delta n}\equiv (\delta n_{\rm b},\delta n_{\rm e}, \delta n_{\rm \mu})$. 
In the inner core, where muons are present, these vectors are related as
\begin{eqnarray}
\pmb \eta= \textbf{\textsf{A}}_{(\mu)\,ji} \pmb {\delta n}, \label{eqA}
\end{eqnarray}
where the elements of the first 
row 
in the matrix $\textbf{\textsf{A}}_{(\mu)}$ equal 
$\textbf{\textsf{A}}_{(\mu)\,{\rm n}i}=\partial \delta \mu_{\rm n}/\partial n_i$, 
while the elements of the second and third 
rows
equal $\textbf{\textsf{A}}_{(\mu)\,li}=\partial \eta_{\rm l}/\partial n_i$; 
indices $l$ and $i$ run over $\rm{e,\mu}$ and $\rm{b,e,\mu}$, respectively 
(note that, due to the quasineutrality condition,
$n_{\rm p}=n_{\rm e}+n_{\mu}$, 
any thermodynamic quantity can be presented as a function of three number densities in the $npe\mu$ core). 
We multiply equation (\ref{eqA}) by the matrix 
$\textbf{\textsf{A}}^{-1}_{(\mu)\,ij}$, 
which is the inverse 
to $\textbf{\textsf{A}}_{(\mu)\,ji}$, 
and integrate the result over the volume $V_\mu$, 
where muons are present:
\begin{eqnarray}
\left(\int_{V_\mu}{\rm e}^{-\nu/2}\textbf{\textsf{A}}^{-1}_{(\mu)\,ij}\,dV\right) {\rm e}^{\nu/2}\pmb \eta = \int_{V_\mu} \pmb {\delta n} \,dV. \label{eqmu}
\end{eqnarray}
Here we factored out 
the uniform vector ${\rm e}^{\nu/2} \pmb \eta$.
Then we follow the analogous procedure in the outer core, where muons are absent, and find: 
\begin{eqnarray}
\left(\int_{V_{\rm e}}{\rm e}^{-\nu/2}\textbf{\textsf{A}}^{-1}_{({\rm e})\,ij}\,dV\right) {\rm e}^{\nu/2} \pmb \eta = \int_{V_{\rm e}} \pmb {\delta n}\, dV, \label{eqe1}
\end{eqnarray}
where $\textbf{\textsf{A}}^{-1}_{({\rm e})\,ij}=0$ for the last 
row
and the last column;
integration goes over the outer core, where muons are absent.
Then we sum up equations (\ref{eqmu}) and (\ref{eqe1}) 
to obtain
\begin{eqnarray}
\left(\int_{V_\mu}{\rm e}^{-\nu/2}\textbf{\textsf{A}}^{-1}_{(\mu)\,ij}\,dV+\int_{V_{\rm e}}{\rm e}^{-\nu/2}\textbf{\textsf{A}}^{-1}_{({\rm e})\,ij}\,dV\right) {\rm e}^{\nu/2} \pmb \eta = \nonumber \\
=(\delta N_{\rm b},\delta N_{\rm e},\delta N_{\rm \mu}), 
\label{sum}
\end{eqnarray}
where $\delta N_j=N_j-N_j^{\rm eq}$ 
is the difference between the actual particle number $N_j$ of species $j$ in the core and their equilibrium value, 
$N_j^{\rm eq}$.
In equilibrium both $\pmb \eta$ and $\delta N_j$ 
must 
vanish.
Multiplying equation (\ref{sum}) by the matrix $\textbf{\textsf{G}}_{ji}$, 
which is the inverse 
to the matrix 
$\int_{V_\mu}{\rm e}^{-\nu/2}\textbf{\textsf{A}}^{-1}_{(\mu)\,ij}\,dV+\int_{V_{\rm e}}{\rm e}^{-\nu/2}\textbf{\textsf{A}}^{-1}_{({\rm e})\,ij}\,dV$, 
we find (see \citealt{fr05}):
\begin{eqnarray}
\eta_{\rm l}^\infty=\textbf{\textsf{G}}_{\rm l b} \delta N_{\rm b}+ \textbf{\textsf{G}}_{\rm l e} \delta N_{\rm e}+\textbf{\textsf{G}}_{\rm l \mu} \delta N_{\rm \mu},
\label{eta1}
\end{eqnarray}
where $\rm{l=e,\mu}$.
Taking time derivative, equation (\ref{eta1}) can be presented as
\begin{eqnarray}
\dot{\eta}_{\rm l}^\infty=\textbf{\textsf{G}}_{\rm l b} \delta \dot{N}_{\rm b}+\textbf{\textsf{G}}_{\rm l e} \delta \dot{N}_{\rm e}+\textbf{\textsf{G}}_{\rm l \mu} \delta \dot{N}_{\rm \mu}. \label{etadot}
\end{eqnarray}
where $\delta \dot{N}_i=\dot{N}_i- \dot{N}_i^{\rm eq}$.
The equilibrium number of particle species $i$ in the core, 
$N_i^{\rm eq}$, 
changes due to NS compression in the course of the accretion and/or NS spin-down.
$\dot{N}_i^{\rm eq}$ is responsible 
for 
building up 
the imbalances in the core. 
We determine $\dot{N}_i^{\rm eq}$ following \cite{hartle67,ht68,fr05} 
with all the necessary modifications caused by accretion.
At the same time, 
an actual (non-equilibrium) number of particle species $i$ in the core, $N_i$, 
varies because of the two processes. 
The first process is the particle mutual transformations 
due to
non-equilibrium MUrca reactions. 
These transformations tend to relax the imbalances $\eta_{\rm l}$ generated by the accretion 
and/or spin-down of the star.
The second one is the transformation of the bottom layers of the crust into the homogeneous matter of the core
under compression.
Since this transformation is a rather model-dependent process, 
in what follows we make 
a simplifying assumption 
that the number of baryons in the crust does not change with time. 
In other words, the accretion of $\delta A$ baryons increases the number 
of baryons in the core by the same value, $\delta A$. 
Obviously, $\delta N_{\rm b}=0$ in such a formulation of the problem.
Moreover, we follow \cite{gc20,gc21} who suggested that the transformation of the crust matter into the core matter under compression is warranted by a specific instability that disintegrates nuclei in the inner layers of the crust into 
neutrons (with no admixture of protons and electrons).
In such a paradigm, we have 
\begin{eqnarray}
\dot{N}_{\rm l}=\int_{\rm core} {\rm e}^{\nu/2}\Delta\Gamma_{\rm l}\, dV. \label{Ndot}
\end{eqnarray}
Note that, our conclusions are not really sensitive to the above assumptions about the physics near the crust-core boundary. 
We checked this by considering the two other limiting possible cases: 
(i) the accretion does not change the baryon number in the core; 
(ii) the accretion of $\delta A$ baryons increases the number of baryons in the core by $\delta A$, 
but the nuclei in the inner layers of the crust 
disintegrate to beta-equilibrated homogeneous  
npe-matter of the core 
(not just neutrons).
The cases (i) and (ii) lead 
to equivalent values of $\delta \dot{N}_i$ ($i=\rm b,e,\mu$), and thus to
a similar
evolution of $\eta_{\rm l}$, $T^\infty$, and $T_{\rm s}^\infty$
that only 
slightly
differs from that presented below in this paper
(see Figs.\ \ref{Fig:2}, \ref{Fig:3}, \ref{Fig:TscSC}, and \ref{Fig:TmutSC}).

\section{Physics input}
\label{input}

To illustrate our results, we employ
BSk24 EOS of the BSk (Brussels-Skyrme) family
(\citealt{gcp13, pfcpg13,pcp18,pcp19}). 
This EOS allows for muons in the inner layers of an NS.
In what follows, 
we consider
$1.4M_\odot$ NS model, for which DUrca processes 
are closed in the whole NS core 
(for BSk24 EOS DUrca processes operate in stars with $M>1.595M_\odot$, 
where they accelerate the relaxation of the imbalances, see Section \ref{assumptions}). 

To calculate the neutrino emissivity, $Q_{\nu}$, due to baryon-baryon bremsstrahlung and non-equilibrium MUrca processes, 
as well as the net reaction rates, $\Delta \Gamma_{\rm l}$, which appear in the
evolution equations of 
Section \ref{approach}, we follow the 
review article by \cite{ykgh01}, 
but account, 
in addition, for the enhancement of MUrca reaction rates reported by \cite{sbh18}. 
More precisely,
we use the formulas from \cite{ykgh01} 
with the equilibrium emissivities for MUrca processes
enhanced by a factor given by the equation (14) of \cite{sbh18}. 
To calculate the heat capacity appearing in equation (\ref{Teq}), we follow \cite{yls99}.

To simplify analysis, in the paper
we assume that neutrons are nonsuperfluid, 
and discuss the effect of neutron superfluidity in Section \ref{assumptions}, arguing that neutron superfluidity does not change the conclusions of the paper qualitatively, since neutron energy gap is expected to be noticeably lower than the 
proton gap. 

\begin{figure}
    \begin{center}
        \leavevmode
        \center{\includegraphics[width=0.77\linewidth]{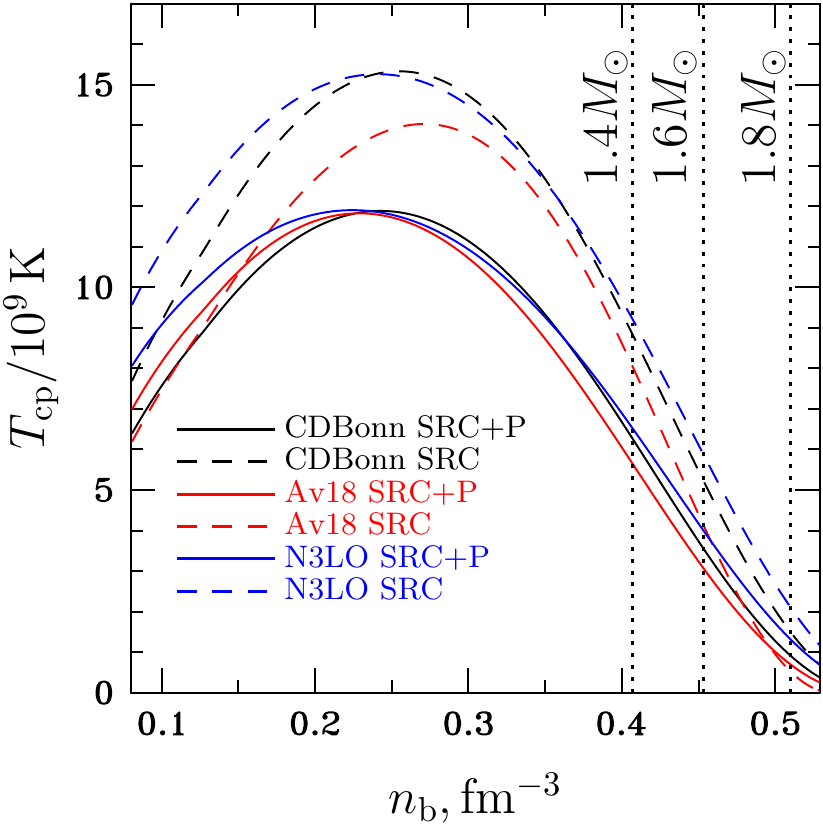}}
    \end{center}
    \caption{
     Local proton critical temperature $T_{\rm cp}$ in the NS core as a function of $n_{\rm b}$ for six models of Ding et al. (2016) (see their table I and notations there). Vertical dotted lines indicate
     central baryon number densities for NSs with $M=1.4M_\odot,1.6M_\odot$, and $1.8M_\odot$.
 }
    \label{Fig:Tc}
\end{figure}

We allow 
for the superconductivity of protons in the NS core and assume that, if present, proton superconductivity occupies the whole core. Moreover, in Section \ref{SC}, 
when calculating the thermal states of an MSP with vanishing magnetic field, 
we, for simplicity, 
assume that the redshifted proton critical temperature is constant throughout the core 
and equals $T_{\rm cp}^\infty=2\times 10^9\rm \, K$. 
Note that microscopic calculations predict 
rather wide profiles of $T_{\rm cp}$, see, e.g., 
Fig.\ \ref{Fig:Tc}, where we demonstrate six critical temperature profiles 
from the paper by
\cite{ding16}, as functions of the baryon number density, $n_{\rm b}$.%
%
\footnote{Actually, \cite{ding16} presented $T_{\rm cp}$ as a function of the proton Fermi momentum, $k_{\rm Fp}$. 
To obtain $T_{\rm cp}$ as a function of $n_{\rm b}$ we make use of the BSk24 EOS.}
Fig.\ \ref{Fig:Tc} shows {\it local} (unredshifted) critical temperatures, 
which are independent of the stellar model.
One can see that proton superconductivity extends over the whole core for NSs of moderate masses
and $T_{\rm cp}$ is, generally, larger than $2\times 10^9\rm \, K$.
Thus, $T_{\rm cp}^\infty=2\times 10^9\rm \, K$ can be considered as a lower limit 
on 
the real $T_{\rm cp}^\infty$ for a chosen model of an NS with $M= 1.4 M_\odot$. 
Below we comment on how our results modify if real minimum of $T_{\rm cp}^\infty$ differs from $2\times 10^9\rm \, K$.
Proton superconductivity 
with $T_{\rm cp}\ga 2\times 10^9$~K 
almost completely 
suppresses the proton heat capacity and proton-baryon bremsstrahlung processes (\citealt{yls99}).
Thus, we assume that, if protons are superconducting, 
they do not contribute to the heat capacity $C$;
the contribution to $C$ from other particle species
remains unaffected, while proton-baryon bremsstrahlung is completely 
suppressed.   
In turn, the reduction factors describing suppression of the non-equilibrium MUrca reactions by proton superconductivity
are, generally, rather cumbersome and have been analyzed in detail by \cite{vh05}.
Fortunately, in the limit $T\ll T_{\rm cp}$ they can be substantially simplified. 
First, in this limit the nonequilibrium MUrca reactions only proceed 
if $|\eta_{\rm l}|>\Delta_{\rm p}$ in the case of neutron branch of MUrca and 
if $|\eta_{\rm l}|>3\Delta_{\rm p}$ in the case of proton branch, 
where $\Delta_{\rm p}$ is the proton energy gap (\citealt{reisenegger97,pr10}).
\cite{pr10} proposed simple analytic expressions 
for the reduction factors of MUrca non-equilibrium reactions 
(see their formulas 20 and 21, and note that the corresponding expression B.15 in their appendix, which is equivalent to equations 20 and 21, contains a typo). 
These expressions are valid for $\Delta_{\rm p}\ga 30k_{\rm B}T$ 
($k_{\rm B}$ is the Stefan-Boltzmann constant).
The above inequality is satisfied in the parameter range of our interest 
(e.g., for $T=10^8\,\rm K$ and $T_{\rm cp}=2\times 10^9\,\rm K$ one has $\Delta_{\rm p}\approx 35k_{\rm B}T$). 
We use these analytic expressions in our calculations.

When considering superconducting NS cores we, 
for the first time in this context, account for the magnetic field existing there.
Depending on the density and microphysics model, 
protons in the NS core form a superconductor 
of either type I or type II (\citealt{gas11,hs18}).%
%
\footnote{Other, more exotic phases are also possible, see an interesting recent work (\citealt*{wgn20}) in this direction.}
%
If protons form a type II superconductor, 
the magnetic field is confined to the flux tubes (Abrikosov vortices) in the core; 
in the case of type I superconductor the field is contained in the 
macroscopic (but small) domains
surrounded by the magnetic-field-free superconducting matter (\citealt{degennes67}).
Protons are normal (nonsuperconducting) both inside the flux tubes 
(in the case of type-II superconductor) and inside the domains (in case of type-I superconductor)
and we shall treat particle mutual transformations as fully unsuppressed there 
(see, e.g., \citealt{sww98}, 
who used this approximation to describe MUrca reactions with localized proton excitations
in the cores of flux tubes).
We expect that this approximation gives correct order-of-magnitude estimates 
for the nonequilibrium MUrca reaction rates
because the wavelengths of neutron, proton, and lepton quasiparticles in NS interiors
are generally smaller than the 
typical sizes of the flux-tube cores and normal domains (e.g., \citealt{gusakov19,dg17}).
Summarizing, we model the NS matter inside the flux tubes/normal domains 
as a uniform non-superconducting liquid.
Note that, for simplicity, we neglect enhancement of the reaction rates 
by the magnetic field 
studied
by \cite{by99},
because this enhancement is quite moderate for the magnetic field strengths 
reached inside the flux tubes/normal domains.

\begin{figure}
	\begin{center}
		\leavevmode
       \includegraphics[width=0.77\linewidth]{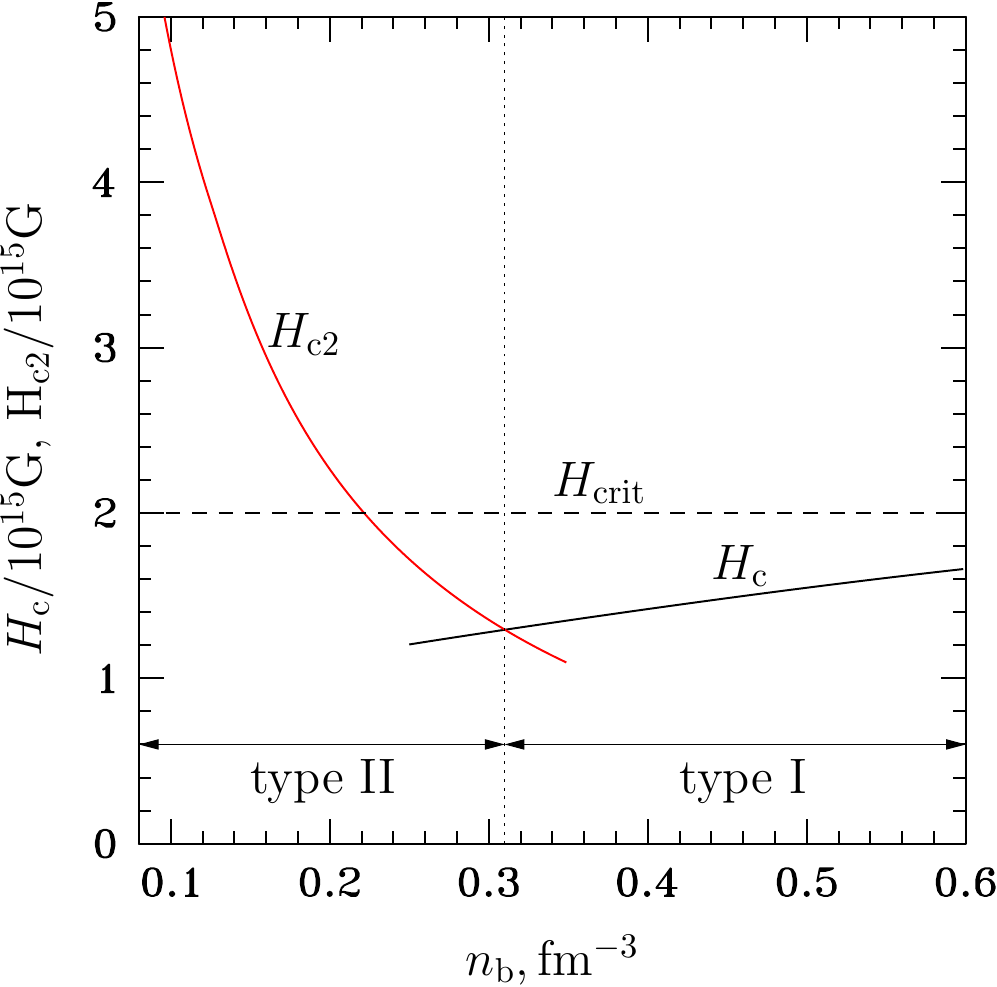}
	\end{center}
	\caption{
	$H_{\rm c}$ and $H_{\rm c2}$ in the NS core as functions of $n_{\rm b}$ for $T_{\rm cp}=2\times 10^9\,\rm K$. Vertical dots show $n_{\rm b}$ corresponding to the interface between type I and type II superconductors. Dashes show adopted in the paper value of $H_{\rm crit}$.
	}
	\label{Fig:Hcrit}
\end{figure}

Clearly, the volume fraction occupied by 
the
nonsuperconducting matter depends on the magnetic field strength in the core, $B$, and can be estimated as $\sim B/H_{\rm crit}$, where $H_{\rm crit}$ is the magnetic field value inside the flux tube/normal domain. 
In the case of type-II superconductor, $H_{\rm crit}$ corresponds 
to the upper critical magnetic field, $H_{\rm c2}$, 
while in type-I superconductor $H_{\rm crit}=H_{\rm c}$ 
(\citealt{degennes67}).
Both critical fields 
$H_{\rm c2}$ and $H_{\rm c}$ vary throughout the core 
by a factor of few
(\citealt{gas11,lp80}), see Fig.\ \ref{Fig:Hcrit}, 
and we adopt their typical value, $\sim 2\times 10^{15}\,\rm G$, as $H_{\rm crit}$ in our calculations.
In what follows, 
both type-I and type-II superconductors are treated as described above, 
i.e., not discriminating
between these two phases. 
Since the value of the magnetic field in the core is highly uncertain 
(see, e.g., \citealt*{crt19} and references therein) 
and can strongly deviate from the value of the NS dipole magnetic field, $B_{\rm dip}$, 
we consider $B$ as a free parameter in the calculations below.

\section{Thermal states of superconducting magnetized MSPs}
\label{SC}

\begin{figure*}
	\begin{center}
		\leavevmode
		\center{\includegraphics[width=0.9\linewidth]{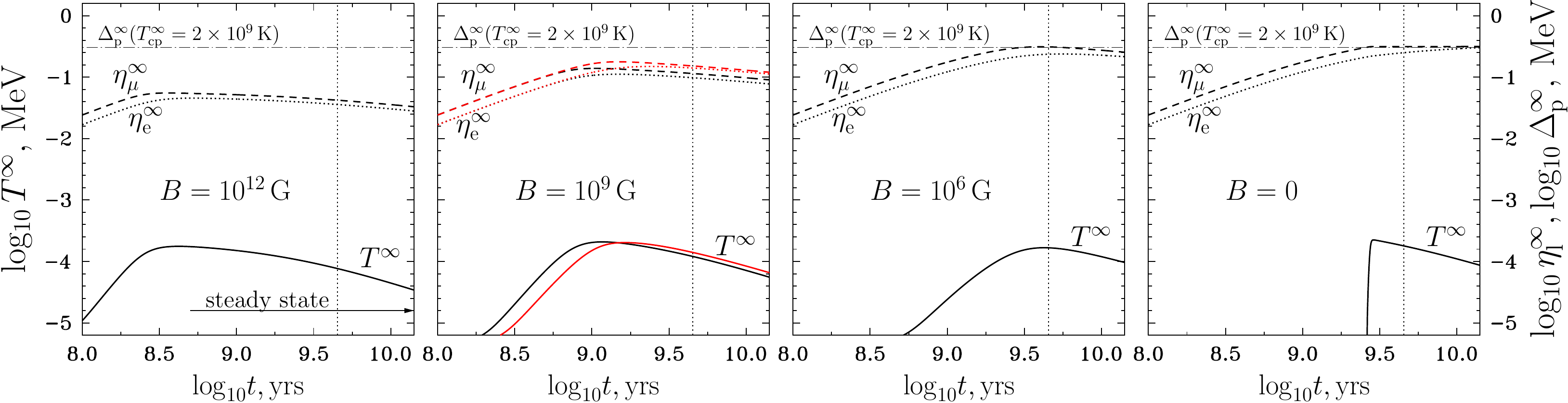}}
	\end{center}
	\caption{
		Internal stellar temperature $T^\infty$ (solid lines)
		and imbalances $\eta_{\rm e}^\infty$ and $\eta_{\rm \mu}^\infty$ (dotted and dashed lines)
		versus time. Black lines account for the enhancement of MUrca reactions 
		as discussed in 
		Section \ref{input}, 
		red lines (in the second panel from the left) disregard this effect. Vertical dots mark 
		the time when $\nu\approx 173\,\rm Hz$ (the spin frequency of J0437).
		Horizontal dot-dashed lines show the redshifted proton superfluid gap, corresponding to $T_{\rm cp}^\infty=2\times 10^9\,\rm K$. At $t=0$ the star is cold and in chemical equilibrium; $\nu(t=0)=300\,\rm Hz$; $B_{\rm dip}=1.6\times 10^8\,\rm G$. 
		We do not show the plot at $t<10^8\,\rm yrs$; 
		it corresponds to the gradual growth of the imbalances.
	}
	\label{Fig:SpinDownB}
\end{figure*}

Here our aim is to analyze the role of the core 
magnetic field 
in the rotochemical heating of superconducting MSPs.
But first, let us assume for a moment
that the magnetic field is absent, 
protons are strongly superconducting in the whole NS core 
($T^\infty\ll T_{\rm cp}^\infty$), $T_{\rm cp}$-profile is flat ($ T_{\rm cp}^\infty=\rm const$), while neutrons are normal.
Since the chemical imbalances $\eta_{\rm l}$ 
relax through the MUrca processes only, 
$\eta_{\rm l}$
will grow until $|\eta_{\rm l}^\infty|>\Delta_{\rm p}^\infty$, 
when the reactions of the MUrca neutron branch are open. 
After this condition is fulfilled for one of the imbalances, 
the reactions will prevent the subsequent growth of the corresponding $\eta_{\rm l}^\infty$
by compensating the effect of compression (\citealt{reisenegger97}).
As a result, $|\eta_{\rm l}^\infty|$ freezes 
at some equilibrium value slightly exceeding $\Delta_{\rm p}^\infty$ 
(see the dashed line in the last panel of Fig.\ \ref{Fig:SpinDownB} at ${\rm log}_{10}t>9.5$). 
Even small variation of $\eta_l^\infty$ results in a strong variation 
of the reaction rates tending to bring $\eta_{\rm l}^\infty$ back to its equilibrium value.  
In what follows, we shall call such quasiequilibrium the `steady state'.
Higher values of $\Delta_{\rm p}^\infty$ correspond to higher values 
of $|\eta_{\rm l}^\infty|$ in the steady state 
and higher energy release in the non-equilibrium reactions. 
Indeed, we can write: 
$\int_V Q_{\rm heat}^\infty dV=\sum_l \eta_{\rm l}^\infty \int_V \Delta \Gamma_{\rm l} {\rm e}^{\nu/2} dV=\sum_{\rm l} \eta_{\rm l}^\infty \dot{N}_{\rm l}$ (see equations \ref{heat} and \ref{Ndot}). 
To calculate $\dot{N}_{\rm l}$ let us consider, for example, 
a situation when $\eta_{\rm \mu}^\infty$ 
have reached the steady-state, 
while $|\eta_{\rm e}^\infty|<\Delta_{\rm p}^\infty$ 
(exactly this situation is realized in the fourth panel 
of Fig.\ \ref{Fig:SpinDownB}). 
Then $\dot{N}_e=0$, since reactions with electrons are locked
(we do not account for the lepton decay here, and discuss its role in Section \ref{assumptions}),
while the quasiequilibrium condition 
$\dot{\eta}_\mu^\infty=0$ prescribes 
$\dot{N}_\mu=\dot{N}_\mu^{\rm eq}+\textbf{\textsf{G}}_{\rm \mu e}/\textbf{\textsf{G}}_{\rm \mu \mu} \dot{N}_{\rm e}^{\rm eq}$ 
(see equation \ref{etadot}). 
Thus, in the steady state $\dot{N}_{\rm l}$ 
are driven by the compression rate and do not depend on the imbalances, 
while the heating rate appears to be proportional to the value 
of the imbalance in the steady state, 
or (approximately)
to $\Delta_{\rm p}^\infty$. 
This property was used by \cite{reisenegger97,pr10,reisenegger15} 
to explain high thermal luminosity of J0437.

\begin{figure*}
	\begin{center}
		\leavevmode
		\center{\includegraphics[width=0.9\linewidth]{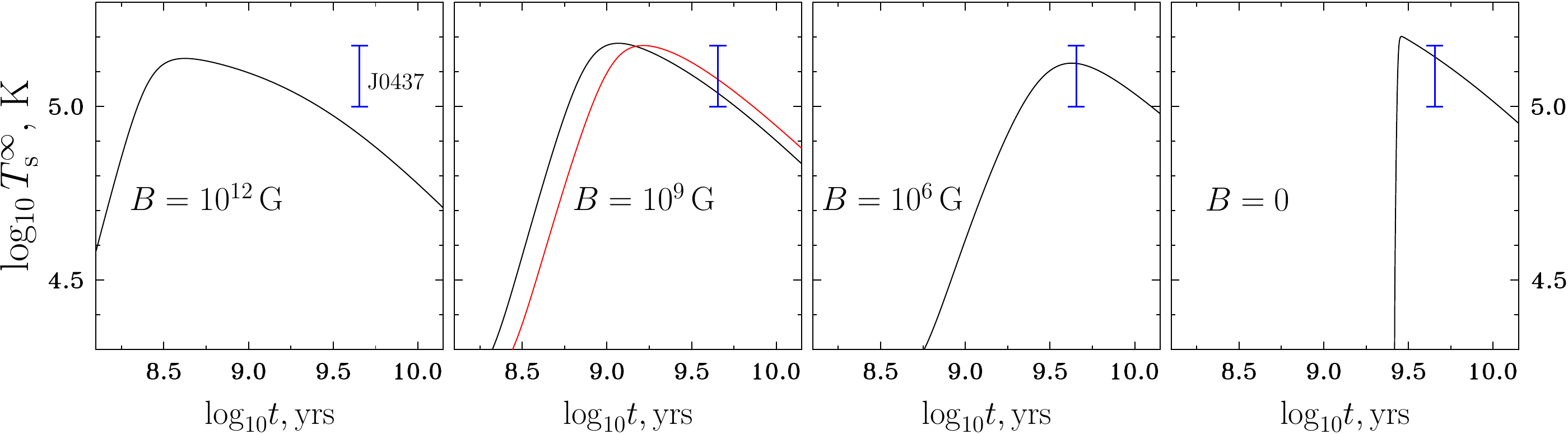}}
	\end{center}
	\caption{
		Surface stellar temperature $T_{\rm s}^\infty$ versus time for the same parameters as in Fig.\ \ref{Fig:SpinDownB}. Notations are the same. 
		Error bars show $T_{\rm s}^\infty$ for the pulsar J0437.
	}
	\label{Fig:Ts}
\end{figure*}

Such an analysis, however, is valid only in the absence of the magnetic field. 
As we already discussed above, the magnetic field makes 
part of the stellar core 
with the volume fraction $\sim B/H_{\rm crit}$ 
nonsuperconducting and allows for unsuppressed nonequilibrium reactions there. 
These reactions may efficiently relax the imbalances
and prevent $|\eta_{\rm l}^\infty|$ from growing to $\Delta_{\rm p}^\infty$.

To analyze the role of the magnetic field,
we model the rotochemical heating of an MSP 
for different values of the core magnetic field, $B$. 
We assume that 
MUrca
reactions 
are not suppressed inside the flux tubes (or normal domains), 
and are \underline{completely forbidden outside} (\citealt{sww98}). 
This is a good approximation, even for small $B$, as long as 
$|\eta_{\rm l}^\infty|<\Delta_{\rm p}^\infty$ and $T^\infty \ll \Delta_{\rm p}^\infty$. 
For example, the reduction factor for 
the neutron branch of MUrca process 
is $\sim 2\times 10^{-11}$ if we take 
$T^\infty=5\times 10^7\,\rm K$, 
$T_{\rm cp}^\infty=2\times 10^9\,\rm K$, and 
$\eta_{\rm l}^\infty=0.8 \Delta_{\rm p}^\infty$. 
This value is much smaller than 
the nonsuperconducting volume fraction,
$B/H_{\rm crit}$, for the values of 
$B$
considered in the numerical examples below.

In this Section, 
we ignore the prehistory of the pulsar, 
namely, its compression at the LMXB stage. 
We assume that, initially 
(at the moment of time $t=0$),
NS 
was in chemical equilibrium,
and rotated at a spin frequency
$\nu(t=0)=\Omega(t=0)/(2{\rm \pi})=300\,\rm Hz$;
the NS temperature  $T^\infty(t=0)$ 
is assumed to be low. 
We also assume that the pulsar spins down due to magneto-dipole losses 
with the spin-down rate $\dot{\Omega}$ equal to
\begin{eqnarray}
	\dot{\Omega}=-\frac{2B_{\rm dip}^2 R^6\Omega^3}{3c^3I}, \label{sd}
\end{eqnarray}
where $I$ is the NS moment of inertia and $B_{\rm dip}=1.6\times 10^8\,\rm G$.%
%
\footnote{Such value of $B_{\rm dip}$ 
corresponds to the actual (i.e., accounting for the Shklovskii effect) 
spin-down rate of J0437.} 
%
While equation (\ref{sd}) describes the energy losses of a rotating dipole in vacuum, 
which is a rather crude emission model, 
its accuracy is quite sufficient for our analysis.

Figure \ref{Fig:SpinDownB} shows the evolution of $T^\infty$, $\eta_{\rm e}^\infty$, 
and $\eta_{\rm \mu}^\infty$ for four values 
of the core magnetic field $B$ ($B=10^{12},10^9,10^6,0\,\rm G$). In case of $B=0$ we assume $T_{\rm cp}^\infty=2\times 10^9\,\rm K$.
One can see that, in the beginning, in all the panels 
chemical imbalances grow with the same rate. 
This rate is 
defined 
by the compression rate of the stellar matter; 
non-equilibrium reactions, which depend on the value of the imbalances, 
are still too 
slow 
to contribute to the evolution of the imbalances. 
Growth of the imbalances $\eta_{\rm l}$ leads to the growth of the non-equilibrium reaction rates. 
As a result, at some moment non-equilibrium reactions 
come into play and start to compensate the effect of compression.
The imbalances reach the steady state when this compensation becomes exact. 
After that, the imbalances stay in the steady state 
(see the arrow in the left panel). 
In the absence of the evolution of the compression rate (${\Omega\dot{\Omega}}=\rm const$),
they would stay exactly constant. 
However, the compression rate decreases 
with time ($\Omega\dot{\Omega}$ decreases, see equation \ref{sd})
and the imbalances trace this decrease. 
Higher values of $B$ result 
in the lower steady-state imbalances and lower $T^\infty$. 
This is not surprising, 
since the higher $B$, 
the larger is the volume 
fraction, where nonequilibrium reactions are unsuppressed 
and can effectively 
relax $\eta_{\rm l}^\infty$. 
Note that the above consideration with non-zero $B$ 
is valid only 
as long as 
$\eta_{\rm l}^\infty<\Delta_{\rm p}^\infty$.
Once $\eta_{\rm l}^\infty$ reaches  $\Delta_{\rm p}^\infty$, 
the reactions start to operate in the whole core 
and this
stops further growth of $\eta_{\rm l}^\infty$. 
For the reference we indicate $\Delta_{\rm p}^\infty$ 
for $T_{\rm cp}^\infty=2\times 10^9\,\rm K$ by dot-dashed line in Fig.\ \ref{Fig:SpinDownB}.
One sees that imbalances do not reach 
$\Delta_{\rm p}^\infty$ for $B\ga 10^6\,\rm G$ 
and assumed spin-down parameters. 
In other words, the magnetic field $B\ga 10^6\,\rm G$ 
reduces the efficiency of rotochemical mechanism. 
Figure \ref{Fig:Ts} illustrates this point,
explicitly showing the corresponding redshifted 
effective 
surface temperature, $T_{\rm s}^\infty$, 
for the same spin-down parameters and the same set of $B$ values as in Fig.\ \ref{Fig:SpinDownB}. 
In Fig.\ \ref{Fig:Ts}
we also show the error bars for the measured 
redshifted effective surface temperature of J0437. 
We choose the horizontal coordinate corresponding 
to the actual spin rate of 
this MSP. 
The time coordinate is also in agreement with the estimated age of J0437,
see \cite{durant12,gr10}.
One can see that at $B\ga 10^{12}\,\rm G$ the rotochemical mechanism becomes unable to explain the observed temperature of J0437 for the adopted stellar model.

In the second panels of Figs.\ \ref{Fig:SpinDownB} and \ref{Fig:Ts} 
we illustrate the effect of the enhancement of MUrca processes 
discussed in \cite{sbh18}. 
For comparison, by red lines we show the results 
obtained neglecting the enhancement of MUrca reactions. 
One can see that, while the quantitative effect is obvious 
(red dashed and dotted lines
pass higher in the steady state, 
because the reactions are not so effective), 
the picture does not change qualitatively, 
and thus our conclusions are not really sensitive 
to the accounting for
the enhancement coefficients. 

We should emphasize that, 
if the proton critical temperature profile does not extend over the whole NS core, 
then the unsuppressed reactions will proceed in the normal part of the core,
which makes the 
role of the magnetic field in establishing the equilibrium
negligible:
Even at $B=0$ the imbalances will relax efficiently.

\section{Effect of preceding accretion on the thermal states of MSPs}
\label{acc}

Let us now analyze how the prehistory of an MSP, 
namely the accretion at the LMXB stage, affects its subsequent thermal 
evolution.

\begin{figure}
	\begin{center}
		\leavevmode
		\includegraphics[width=0.9\linewidth]{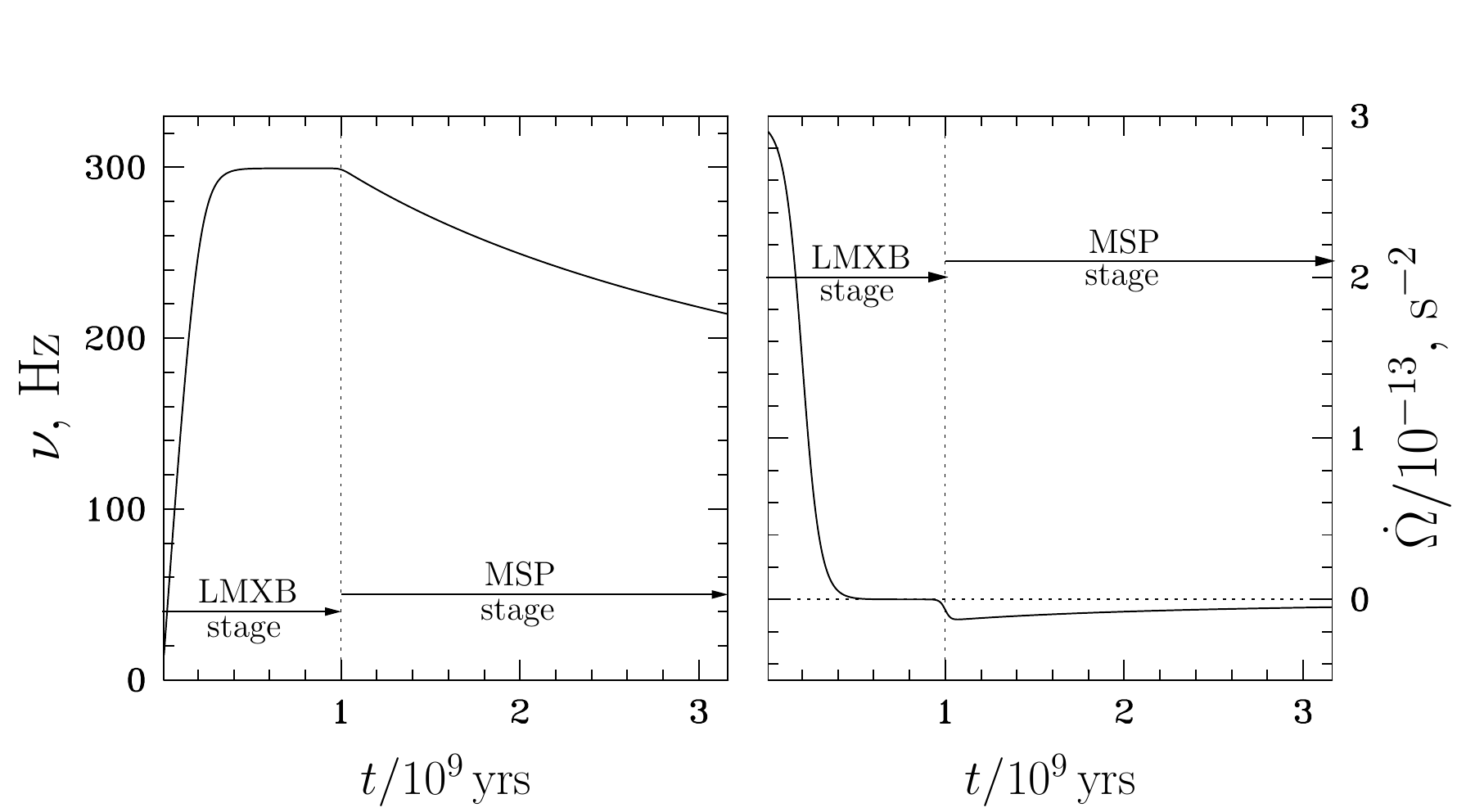}
	\end{center}
	\caption{
		$\nu$ (left panel) and $\dot\Omega$ (right panel) versus time. 
	}
	\label{Fig:1}
\end{figure}

We assume that the accretion proceeds with constant 
average rate $\dot{M}$ during some period of time $t_{\rm acc}$ 
(in our numerical examples $\dot{M}=10^{-10}M_\odot/\rm yr$, 
$t_{\rm acc}=10^9\,\rm yrs$) 
and then smoothly 
(on a timescale $\sim 3\times 10^7\,\rm yrs$, \citealt{tauris12}) 
switches off. 
We also assume that during almost the whole LMXB phase (except for the beginning and final stages, see below for details), 
when accretion is active, an NS has an equilibrium frequency 
$\nu_{\rm eq}$, 
defined by the condition that the 
accretion spin-up is balanced by the magneto-dipole spin-down.
In our numerical calculations we choose 
$\nu_{\rm eq}=300\, \rm Hz$. 
When the accretion 
smoothly switches off, the star starts to spin down. 
The spin-down rate increases 
gradually
on the time-scale $\sim 3\times 10^7\,\rm yrs$ from zero to the value given by
equation (\ref{sd}).
In the beginning of the accretion phase we take into account
an initial spin up of accreting NS up to the equilibrium frequency $\nu_{\rm eq}$.
We choose the duration of this spin-up stage 
to be $\sim 2\times 10^8\,\rm yrs$, 
in accordance with the observed $\dot{\nu}$ for some NSs in LMXBs during outbursts 
(see, e.g., \citealt{papitto08,patruno10,papitto11}) 
and in accordance with the accretion torque modeling
(\citealt{pr72,gl79}). 
Figure \ref{Fig:1} shows the behavior of $\nu$ and $\dot{\Omega}$ with time.
Then, using equations of Section \ref{approach}, 
we model the joint thermal and chemical evolution of an NS.

\subsection{Nonsuperconducting NS matter}
\label{nonSF}

\begin{figure}
	\begin{center}
		\leavevmode
		\includegraphics[width=0.77\linewidth]{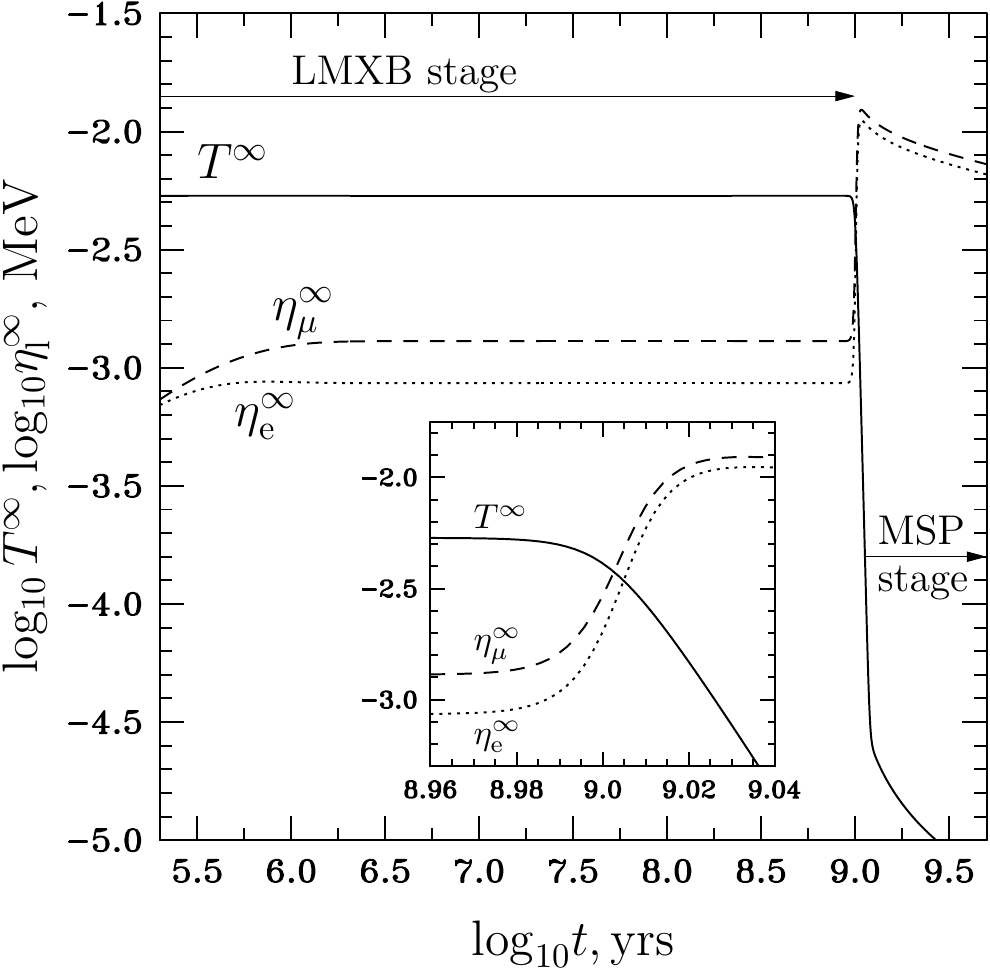}
	\end{center}
	\caption{
		$T^\infty$ (solid line) and imbalances $\eta_{\rm e}^\infty$ (dotted line) and $\eta_\mu^\infty$ (dashed line) versus time.
		To highlight the accretion effect we neglect the 
		effect of $\Omega$-variation on $T^\infty$, $\eta_{\rm e}^\infty$, and $\eta_{\mu}^\infty$.
		We do not show the plot at ${\rm log}_{10} t<5.3$, it corresponds to the gradual growth of the imbalances with time.
	}
	\label{Fig:2}
\end{figure}

\begin{figure}
	\begin{center}
		\leavevmode
		\includegraphics[width=0.77\linewidth]{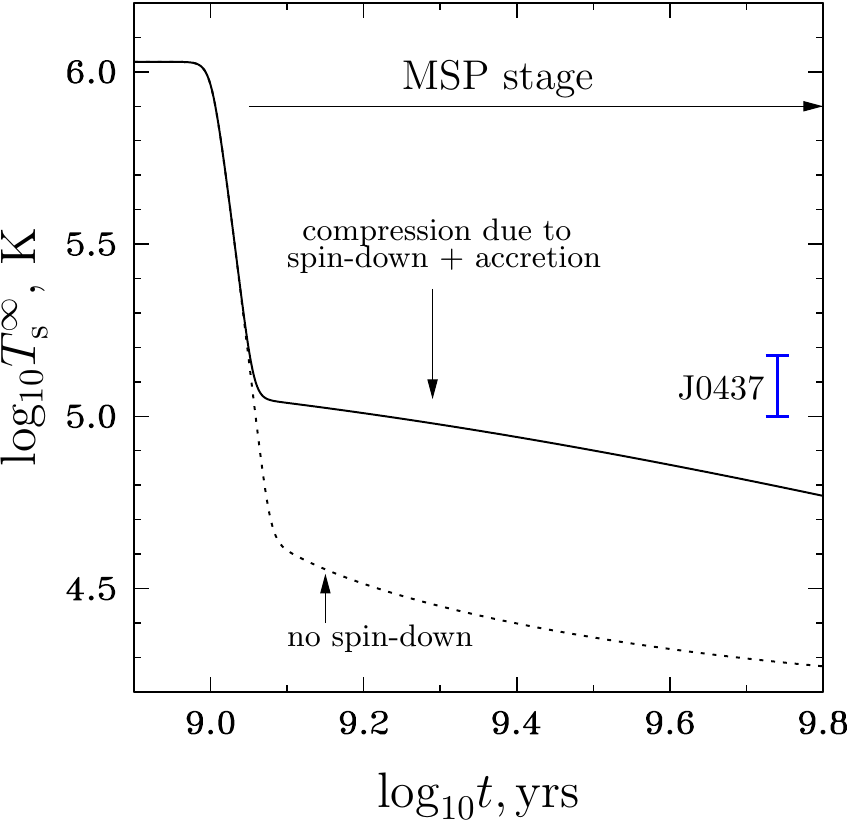}
	\end{center}
	\caption{
		$T_{\rm s}^\infty$ normalized to $10^5$~K
		versus time.
		Dotted line is plotted assuming that there is no spin-down ($\Omega$ is kept constant), 
		solid line accounts for both compression due to accretion (at the LMXB stage) and NS spin-down. 
		The evolution exclusively due to spin-down follows solid line very closely. 
		Error bar shows $T_{\rm s}^\infty$ for the pulsar J0437.
	}
	\label{Fig:3}
\end{figure}

First, we consider a normal (nonsuperconducting) NS. 
Figure \ref{Fig:2} shows temporal evolution of $T^\infty$, $\eta_{\rm e}^\infty$, and $\eta_\mu^\infty$. 
In this figure we neglect $\Omega$ 
evolution to highlight the accretion effect 
(see Fig.\ \ref{Fig:3} for the combined effect of accretion and 
$\Omega$-variation).
The compression of NS material in the course of accretion drives an NS 
out of chemical equilibrium.
As a result, $\eta_{\rm e}^\infty$ and $\eta_\mu^\infty$ grow 
until they reach the equilibrium values 
when particle transformations become fast enough 
to compensate the subsequent growth of $\eta_{\rm e}^\infty$ and $\eta_\mu^\infty$.  
At the LMXB stage the value of $T^\infty$ 
corresponds to the equilibrium one, 
defined by the balance between the NS cooling 
and the deep crustal heating 
(as we already mentioned in Section \ref{approach},
we assume $q=0.5\,\rm MeV$ per accreted nucleon, see \citealt{gc20,gc21}). 
NS heating due to non-equilibrium particle transformations 
proceeding to restore the chemical equilibrium in the core also takes place, 
however, the energy released in these reactions (chemical heating) 
is much smaller than the deep crustal heating.
When accretion ceases at $t\sim 10^9\,\rm yrs$, 
the NS internal temperature $T^\infty$ 
drops on a typical NS cooling timescale. 
However, at some value of $T^\infty$, 
NS cooling becomes approximately balanced by the chemical heating in the core. 
Then the temperature fall
slows down and subsequently (at $t \gtrsim 10^{9.15}$~yrs)
$T^\infty$ evolves
on the timescale of chemical evolution. 
Generally, after the accretion ceases, 
the imbalances are driven by two trends. 
First, $\eta_{\rm e}^\infty$ and $\eta_\mu^\infty$ 
tend to relax to zero by means of particle transformations. 
Second, the NS matter continues to be compressed 
due to NS spin-down that maintains the non-zero imbalances. 
Note, however, that Fig.\ \ref{Fig:2} neglects the spin-down 
($\Omega$ is artificially kept constant) so that in this figure 
the imbalances at the MSP stage are driven by the relaxation through non-equilibrium reactions only.
The reader can notice a sharp increase of $\eta_{\rm e}^\infty$ and $\eta_\mu^\infty$
at $t\sim 10^9\,\rm yrs$, when accretion ceases 
(see also inset in Fig.\ \ref{Fig:2}). 
It is related to the fact that when the accretion rate starts to decrease 
(we remind that this process lasts $\sim 3\times 10^7$~yrs), 
the NS temperature rapidly falls down. 
Driven by strong temperature dependence, the reaction rates decrease as well, 
making the relaxation of the imbalances inefficient. 
As a result, $\eta_{\rm e}^\infty$ and $\eta_\mu^\infty$ get an opportunity to grow up. 
However, at some moment the growth of the imbalances stops.
The main reason for that is that
then the accretion rate becomes too low to provide sufficient compression of NS matter.

Fig.\ \ref{Fig:3} shows the dependence $T_{\rm s}^\infty(t)$. 
The surface temperature traces $T^\infty$ behavior.
Solid line accounts for the compression 
of NS material in the course of both accretion and spin-down ($\Omega$ is now allowed to vary); 
dotted line shows the effect of accretion only ($\dot{\Omega}=0$).
If we account for NS spin-down only, neglecting 
compression of the star
at the LMXB stage 
($t\lesssim 10^9\,\rm yrs$) 
by the accreted material,
we obtain the line that practically coincides with the solid line.
Figure \ref{Fig:3} implies that in the normal NSs compression due to accretion alone (dots) does not generate sufficient imbalances to maintain relatively high NS surface temperatures after the accretion ceases, 
comparable to those observed in some MSPs.
The reason for that is
reactions of particle transformations are fast enough to 
keep the NS core close 
to the equilibrium ($\eta_{\rm e}^\infty\approx \eta_{\rm \mu}^\infty\approx 0$)
despite the strong compression caused by accretion. 
Variation of $q$, $\dot{M}$, as well as  account for $\dot{M}$
temporal dependence
do not change the situation 
qualitatively.

\subsection{Superfluid/superconducting NS matter}
\label{SF}

\begin{figure*}
	\begin{center}
		\leavevmode
		\includegraphics[width=0.9\linewidth]{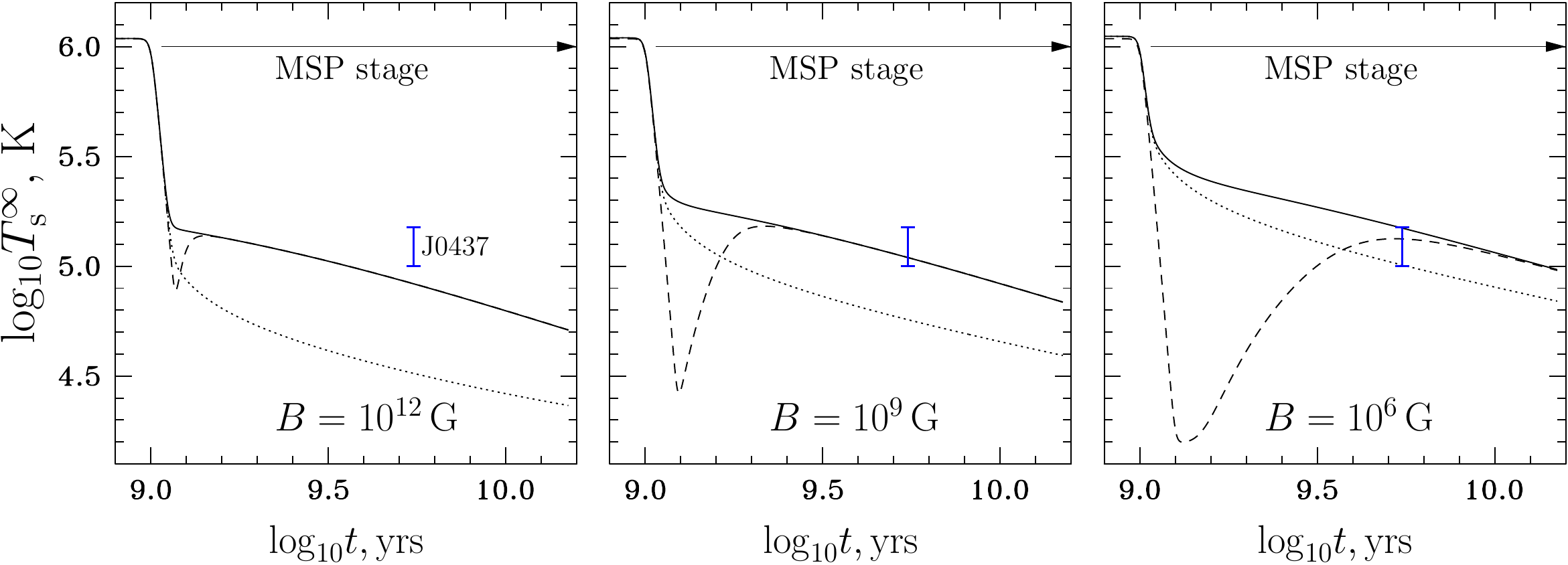}
	\end{center}
	\caption{
		$T_{\rm s}^\infty$ normalized to $10^5$~K
		versus time.
		Protons are superconducting in the whole NS core. Dotted lines are plotted assuming 
		that there is no spin-down ($\dot{\Omega}=0$), 
		dashed lines -- no compression due to accretion at the LMXB stage, 
		solid lines account for both compression due to accretion and NS spin-down. 
		Three panels correspond to three values of the magnetic field (from left to right): 
		$B=10^{12}$, $10^{9}$, and $10^{6}$~G. 
		Error bars show $T_{\rm s}^\infty$ for the pulsar J0437.
	}
	\label{Fig:TscSC}
\end{figure*}

\begin{figure*}
	\begin{center}
		\leavevmode
		\includegraphics[width=0.9\linewidth]{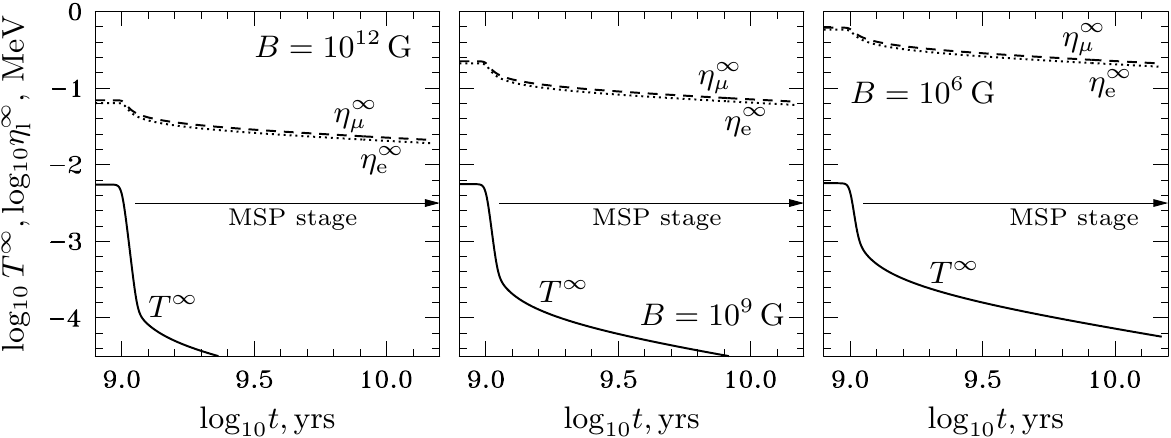}
	\end{center}
	\caption{
		The same as in Fig.\ \ref{Fig:2},		
		but protons are superconducting. 
		It is assumed that $\dot{\Omega}=0$. 
		Three panels correspond to three values of the magnetic field in the core.
	}
	\label{Fig:TmutSC}
\end{figure*}

Baryon superfluidity/superconductivity suppresses the nonequilibrium reaction rates. 
Thus, 
in superfluid/superconducting NSs, 
we can expect that higher values of the imbalances will be reached 
at the LMXB stage,
which can affect the NS surface temperature 
at the subsequent MSP stage.
To illustrate this point, 
let us assume that protons are strongly superconducting, 
while neutrons are normal. 
We allow for non-zero magnetic field in the core (characterized by the parameter $B$)
and model the NS evolution assuming that the proton energy gap 
is high enough, so that the inequality $\eta_{\rm l}^\infty<\Delta_{\rm p}^\infty$ is always satisfied, and MUrca reactions can be treated as fully suppressed outside the flux tubes/normal domains. 
Figure \ref{Fig:TscSC} shows $T_{\rm s}^\infty$ for 
the three values of $B$. 
Lower magnetic fields result in higher $T_{\rm s}^\infty$ 
after 
cessation of accretion
(at $t>10^9\,\rm yrs$). 
This is not surprising, 
since the imbalances
generated in the star during its evolution
are higher for lower $B$, 
see Fig.\ \ref{Fig:TmutSC}. 
\footnote{
\label{foot4}	
Note that, in the case of $B=10^6\,\rm G$ 
(third panel in Figs.\ \ref{Fig:TscSC} and \ref{Fig:TmutSC}) 
the condition $\eta_{\rm l}^\infty<\Delta_{\rm p}^\infty$ assumed in our calculations  
is only
fulfilled for $T_{\rm cp}$-profiles with $T_{\rm cp}^\infty\ga 5\times 10^9\,\rm K$, 
while in the cases $B=10^9\,\rm G$ and $B=10^{12}\,\rm G$ the condition $T_{\rm cp}^\infty\ga 2\times 10^9\,\rm K$ is sufficient.}
At the same time, for 
sufficiently large magnetic fields such as $B=10^{12}\,\rm G$
the generated imbalances are too low to  explain the surface temperature of J0437.
Another interesting feature is that, for MSPs with even vanishing spin-down rate, 
$T_{\rm s}^\infty$ can be relatively high 
during some period of time after accretion switches off 
and NS becomes an MSP. 
For example, for $B=10^9\,\rm G$ the surface temperature remains 
$T_{\rm s}^\infty\ga 10^5\,\rm K$ during approximately $10^9\,\rm yrs$ 
after accretion ceases at 
vanishing
spin-down rate (see dotted line in the middle panel of Fig.\ \ref{Fig:TscSC}). 
In turn, for $B=10^6\,\rm G$ this period lasts approximately $4.5\times 10^9\,\rm yrs$.
During this time $T_{\rm s}^\infty$ is 
supported by relaxation of the imbalances generated during compression of matter by accretion
at the LMXB stage. 
In the case of $B=10^6\,\rm G$ this heating could support the observed temperature of J0437 even in the absence of the pulsar spin-down.

Thus, our results imply that
temperature of an MSP may be directly related to
the evolution of the
star at the LMXB stage.
Millisecond pulsars with low $\Omega \dot{\Omega}$ may nevertheless have rather high surface temperatures ($T_{\rm s}^\infty\ga 10^5\,\rm K$).

As in the case of normal NSs, we
checked the sensitivity of our results to the choice of $\dot{M}$ and $q$. 
We found that, while 
variation of these parameters
has a certain effect on the stellar temperature 
at the LMXB stage, 
the built-up imbalances, 
and hence the NS temperature at the MSP stage, 
are quite insensitive to the choice of $\dot{M}$ and $q$. 
Note that Fig.\ \ref{Fig:TmutSC} 
implies that when we consider superconducting NSs, 
$\eta_{\rm l}^\infty\gg T^\infty$ both at the MSP and LMXB stages 
(i.e., we are in the so-called `suprathermal' regime). 
In this regime, the relaxation rate of the imbalances, 
determined by $\Delta \Gamma_{\rm l}$, 
does not depend on temperature, 
but strongly depends on the imbalance values, 
$\Delta \Gamma_{\rm l}\propto \eta_{\rm l}^7$ (\citealt{ykgh01}). 
As a result, 
although $T^\infty$ is sensitive to $\dot{M}$ and $q$ values, 
the imbalances and NS temperature at the MSP stage are not affected by these parameters.

\section{Discussion}
\label{assumptions}

In this paper we worked under a number of simplifying assumptions. 
Here we discuss how 
they may affect our conclusions.

First, we considered the NS model with forbidden DUrca processes. 
However, for many 
admissible EOSs these processes may operate in the 
inner cores of massive NSs.
If open, DUrca processes dramatically accelerate particle mutual transformations 
and hence relaxation of the imbalances. 
As long as
the magnetic field in NS cores 
does not vanish,
the unsuppressed DUrca reactions proceed in flux tubes/normal domains.
As a result, the steady-states reached in NSs with open DUrca correspond
to much lower values of the imbalances 
comparing to NSs with closed DUrca, 
so that the chemical heating in 
the former stars is not significant.
Note also that the central regions in high-mass NSs
may be non-superconducting
(see, e.g., Fig.\ \ref{Fig:Tc}). 
In these regions particle mutual transformations 
are not suppressed by proton superconductivity, that 
also leads to a more rapid relaxation of the imbalances. 
All this suggests that chemical heating in massive NSs may be inefficient.

Further, in this work we did not account for the non-equilibrium lepton decay process (\citealt{ag10}):
\begin{eqnarray}
	{\rm e}+{\rm l} \rightarrow {\rm \mu} +{\rm l}  +\nu_{\rm e} +\overline{\nu}_{\rm \mu},\,\,\, {\rm \mu}+{\rm l} \rightarrow {\rm e} +{\rm l}  +\nu_{\rm \mu} +\overline{\nu}_{\rm e}. \label{lepton}
\end{eqnarray}
Generally, this process is much weaker than MUrca processes 
(as follows from figure 4 of \citealt{ag10}, by about eight orders of magnitude). 
However, when superfluidity/superconductivity 
strongly
suppresses MUrca reactions, 
lepton decay becomes the main process of particle mutual transformations. 
Note that \cite{ag10} calculated the non-equilibrium reaction rates for (\ref{lepton}) 
in subthermal regime, 
when $\mu_{\rm e}-\mu_{\rm \mu}\ll T$. 
In our problem we are interested in the opposite (suprathermal) regime, 
when $\mu_{\rm e}-\mu_{\rm \mu}\gg T$. 
Unfortunately, lepton decay in this limit has not been discussed in the literature
to our best knowledge, so we did not include this process 
in this work.
However, we believe that lepton decay cannot qualitatively affect our results 
since it relaxes $\mu_{\rm e}-\mu_{\rm \mu}$ only, 
tending to equalize $\eta_{\rm e}$ and $\eta_{\rm \mu}$, 
but it is unable 
to vanish $\eta_{\rm l}$.
Note that, in all our calculations $\eta_{\rm e}$ and $\eta_{\rm \mu}$ 
are rather close to each other,
which means that even if the lepton decay 
was efficient, 
it would equalize $\eta_{\rm e}$ and $\eta_{\rm \mu}$ at some average value, 
but would not significantly affect the evolution of stellar temperature.

Next, for simplicity we assumed that neutrons in the star are nonsuperfluid (normal).
Neutron superfluidity may affect the chemical heating in two ways. 
First, it introduces the Cooper pairing neutrino emission process,
a strong cooling agent in NSs with internal temperatures 
comparable, but smaller than
the neutron critical temperature (\citealt{ykgh01}). 
Note that, due to rather low temperatures of MSPs, 
the Cooper pairing neutrino emission process is negligible in MSPs.
This process
may, however, increase the NS cooling rate at the LMXB stage, 
when an NS is noticeably hotter. 
As a result, the equilibrium temperature at the LMXB stage
may appear to be a bit smaller than in our simulations.

The second effect of neutron superfluidity concerns the suppression of 
particle mutual transformations in MUrca reactions.%
%
\footnote{
	Bremsstrahlung processes with neutrons are also suppressed by neutron superfluidity.}
It directly reduces the cooling/heating rate (depending on the ratio of $\eta_{\rm l}/T$; see \citealt{fr05})
of an NS due to MUrca processes and,
in addition, reduces the relaxation rates of the imbalances, 
allowing 
$\eta_{\rm e}$ and $\eta_{\rm \mu}$
to reach higher values (\citealt*{ynh20}). 
However, such a suppression of MUrca reactions is only efficient 
as long as $\eta_{\rm l}\la \Delta_{\rm n}$, where $\Delta_{\rm n}$ is the neutron superfluid energy gap. 
Once $\eta_{\rm l}$ have reached $\Delta_{\rm n}$, 
the subsequent growth of the imbalances 
leads to less and less efficient suppression 
of the reaction rates by neutron superfluidity. 
In other words, neutron superfluidity 
has a significant  
impact
on the MUrca reaction rates only 
if
$\eta_{\rm l}\la \Delta_{\rm n}$ 
(or $\eta_{\rm l}^\infty \la \Delta_{\rm n}^\infty$) in a substantial fraction of the NS core.
However, observations of cooling NSs 
and NSs in LMXBs
seem to indicate
(e.g., \citealt{gkyg04,plps04,gkyg05,syhhp11,page11,CasA13,CasA15,bhsp18,kgd20,kgd21,sow21})
that 
the maximum value of the redshifted neutron energy gap in the core should not exceed 
$\Delta_{\rm n}^\infty\la 0.07\,\rm MeV$, and is even lower away from the maximum. 
This constraint also
does not contradict microscopic calculations 
(e.g., \citealt{ls01, yls99,gps14,dlz14, drddwcp16, sc19}).
At the same time, 
our results imply that the chemical heating 
is capable of maintaining the surface temperature $T_{\rm s}^\infty\approx 10^5\rm\, K$ 
for the parameters of PSR J0437-4715 as an example, 
only if the imbalances are rather large, 
$\eta_{\rm l}^\infty> 0.1\,\rm MeV$, 
i.e.,
$\eta_{\rm l}^\infty> \Delta_{\rm n}^\infty$. 
This suggests that neutron superfluidity should not affect our results qualitatively, 
although some quantitative effect is 
of course expected.

Now, let us comment on
our model of proton superconductivity.
When considering superconducting NSs, 
we assumed that proton superconductivity 
extends over the whole NS core. 
Only in this case our conclusions 
about the thermal states of superconducting MSPs are valid. 
If $\Delta_{\rm p}\la T$ in some part of the core, 
the imbalances effectively relax 
through unsuppressed MUrca reactions there, 
and no effective chemical heating is possible 
(see \citealt{reisenegger15}).
Moreover, 
to find
the thermal state of a superconducting spinning down MSP 
with $B=0$, 
we assumed that the redshifted proton critical temperature 
is constant throughout the core. 
%
\footnote{
We have to specify the $T_{\rm cp}^\infty$-profile
to calculate the reduction factors for MUrca processes
by proton superconductivity. In the case of non-vanishing magnetic field,
$B\neq 0$, we do not need to calculate these factors, because  
we treat MUrca reactions outside the flux tubes/normal domains as fully suppressed in this case
(i.e., we neglect a contribution of MUrca reactions outside the flux tubes/normal domains in comparison 
to that inside the flux tubes/normal domains).
This approach is justified for typical LMXB/MSP temperatures and magnetic fields $B \gtrsim 10^6$~G
as long as $\eta_{\rm l}^\infty<\Delta_{\rm p}^\infty$ in the whole stellar core.
The latter condition is fulfilled for most of our numerical examples 
for the critical temperature profiles with $T_{\rm cp}^\infty\ga 2\times 10^9\, \rm K$, 
see footnote \ref{foot4} for details.}
In contrast, \cite{reisenegger15} considered a more realistic bell-shaped profile 
of $T_{\rm cp}$. 
They found that the redshifted imbalances 
first grow up to the lowest value of $\Delta_{\rm p}^\infty$ in the core,
then nonequilibrium MUrca reactions switch on 
in the volume fraction in which $\eta_{\rm l}^\infty>\Delta_{\rm p}^\infty$
and prevent subsequent growth of the imbalances in the NS core.
This means that our results for the flat profile 
with $T_{\rm cp}^\infty=2\times 10^9\, \rm K$ 
are equally valid for {\it any} 
bell-shaped profile with the minimum value of 
$T_{\rm cp}^\infty$ equal to $2\times 10^9\, \rm K$.
Note that microscopic calculations 
(see, e.g., \citealt{drddwcp16} and Fig.\ \ref{Fig:Tc}) 
imply that in our stellar model the chosen value of 
$T_{\rm cp}^\infty=2\times 10^9\, \rm K$ can be considered 
as a lower limit 
on the real minimum of $T_{\rm cp}^\infty$. 
Thus, the derived surface temperature 
(see the last panel in Fig.\ \ref{Fig:Ts}) 
is the estimate from below for the real one.

Finally, modeling of the Roche-lobe decoupling phase 
predicts (\citealt{tauris12})  
that in the end of the LMXB phase 
an NS experiences a spin-down torque, 
which 
can
substantially reduce the NS spin frequency. 
Such a spin-down would additionally compress the NS matter.
However, as we checked, 
this effect 
does not significantly affect our results, 
since the matter compression in the course of accretion is much more efficient.

\section{Conclusions}
\label{conc}

This work
studies two aspects of the chemical heating of MSPs. 
First, we analyze the effect of stellar core magnetic field on the heating
of superconducting MSPs.
Second, we look into the role of the preceding accretion at the LMXB stage
on building up the chemical potential imbalances and
their effect on the subsequent thermal evolution of MSP.

In the previous studies of 
\cite{pr10} and \cite{reisenegger15} it was found that an MSP 
should be effectively heated by the rotochemical  
mechanism if
proton and/or neutron energy gaps 
are sufficiently large
in the whole stellar core. 
This is required for an efficient heating since
the redshifted chemical potential imbalances $\eta_{\rm l}^\infty$ tend to grow
to the lowest value of the corresponding redshifted gap
in the course of NS compression. 
Higher values of the gaps 
lead to larger
imbalances, when an NS reaches the steady state. 
Since the heating rate due to the non-equilibrium 
MUrca processes 
is proportional to $\eta_{\rm l}^\infty$ in 
this state, larger
gaps result in hotter NSs in 
the steady state.
Adopting the flat critical temperature profile 
throughout the core, $T_{\rm cp}^\infty=\rm const$,
we confirmed that in the absence 
of the magnetic field the 
imbalances tend to grow 
to the value of the proton superfluid gap ($\eta_{\rm l}^\infty\approx \Delta_{\rm p}^\infty$). 
However, we found that 
even small magnetic field in the core
(in our numerical examples, $B\ga 10^6\,\rm G$) may terminate this growth at some value of $\eta_{\rm l}^\infty<\Delta_{\rm p}^\infty$. 
This happens because the core magnetic field destroys 
superconductivity in some volume fraction 
proportional to the field value,
so that
unsuppressed nonequilibrium reactions may
proceed there
and relax the imbalances. 
As a result, 
the steady state corresponds to lower $\eta_{\rm l}^\infty$ 
and lower surface temperatures at higher $B$. 
Thus, the core magnetic field reduces the
efficiency of the rotochemical heating
even if proton superconductivity extends over the whole NS core.
In particular,
sufficiently high values of the core magnetic field (such as $B=10^{12}\,\rm G$) 
significantly complicate explanation of 
the surface temperature of J0437 within the rotochemical heating scenario.

On the contrary, 
the preceding accretion at the LMXB stage 
may help to bring the imbalances 
to higher values 
even at non-zero magnetic field in the core. 
It appears to be possible
because of the strong compression of NS interiors caused by the accreted matter. 
As a result, after accretion ceases, 
nonequilibrium processes heat an MSP, 
and the latter may stay warm ($T_{\rm s}^\infty\ga 10^5\,\rm K$) for about 
billion years even in the absence of any spin-down (see Fig.\ \ref{Fig:TscSC}). 
In this respect it is worth noting 
that all the known effective reheating mechanisms 
operate by transforming the rotational energy 
to the thermal energy (\citealt{alpar84,reisenegger95,gkr15}). 
Thus, currently, it is believed that the higher is the loss of the rotational energy by an MSP, 
the larger is its surface temperature.
Here we find that it is not necessary the case, 
since preceding accretion may `charge' the MSP core with the chemical energy, 
thus providing the long-lasting energy source for the MSP's subsequent thermal luminosity. 
This means that \underline{MSPs with low spin-down rates may still be warm}.  

Finally, it is worth noting that the magnetic field 
in superconducting cores of ordinary pulsars 
may also reduce the efficiency of rotochemical heating 
in these objects.
We leave a detailed discussion of the related effects for a future work.

\section*{Acknowledgments}
We thank Andreas Reisenegger for valuable comments on the draft version of the manuscript. 
EK was supported by the Russian Science Foundation 
(grant number 19-12-00133).

\section*{Data availability}
The data underlying this article are available in the article.

\label{lastpage}

\end{document}